\documentclass[10pt,conference]{IEEEtran}
\usepackage{algorithm}
\usepackage{algorithmic}
\usepackage{graphicx}
\usepackage{graphics} 
\usepackage{epsfig}

\begin{document}

\title{Generalized Statistics Framework for Rate Distortion Theory with Bregman Divergences}

\author{
\authorblockN{R. C. Venkatesan}
\authorblockA{Systems Research Corporation \\
Aundh, Pune 411007, India \\
ravi@systemsresearchcorp.com} }
%

\maketitle

\begin{abstract}
A variational principle for the rate distortion (RD) theory with
Bregman divergences is formulated within the ambit of the
generalized (nonextensive) statistics of Tsallis. The
Tsallis-Bregman RD lower bound is established. Alternate
minimization schemes for the generalized Bregman RD (GBRD) theory
are derived. A computational strategy to implement the GBRD model is
presented. The efficacy of the GBRD model is exemplified with the
aid of numerical simulations.
\end{abstract}

\section{Introduction}
The generalized (nonextensive) statistics of Tsallis [1,2] has
recently been the focus of much attention in statistical physics,
and allied disciplines \footnote{A continually updated bibliography
of works related to nonextensive statistics may be found at
http://tsallis.cat.cbpf.br/biblio.htm.}. Nonextensive statistics
generalizes the extensive Boltzmann-Gibbs statistics, and has found
much utility in complex systems possessing long range correlations,
fluctuations, ergodicity, chirality and fractal behavior. By definition, the Tsallis entropy is defined in terms of
discrete variables as

\begin{equation}
S_q \left( x \right) = -\frac{{1 - \sum\limits_x {p^q \left( x
\right)} }}{{1 - q}},where,\sum\limits_x {p\left( x \right)}  = 1.
\end{equation}

The constant $ q $ is referred to as the \textit{nonextensivity
parameter}. Given two independent variables $ x $ and $ y $, one of
the fundamental consequences of nonextensivity is demonstrated by
the \textit{pseudo-additivity} relation
\begin{equation}
S_q \left( {xy} \right) = S_q \left( x \right) + S_q \left( y
\right) + \left( {1 - q} \right)S_q \left( x \right)S_q \left( y
\right).
\end{equation}
Here, (1) and (2) imply that extensive statistics is recovered as $
q \to 1 $.  Taking the limit $ q \to 1 $ in (1) and evoking
l'Hospital's rule, $ S_q \left( x \right) \to S\left( x \right) $,
the Shannon entropy. The \textit{jointly convex} generalized
Kullback-Leibler divergence (K-Ld) is of the form [3]

\begin{equation}
I_q \left( {p\left( x \right)\left\| {r\left( x \right)} \right.}
\right) = \sum\limits_x {p\left( x \right)} \frac{{\left(
{\frac{{p\left( x \right)}}{{r\left( x \right)}}} \right)^{q - 1}  -
1}}{{q - 1}}. \
\end{equation}
In the limit $ q \to 1 $, the extensive K-Ld is readily recovered.
Akin to the Tsallis entropy, the generalized K-Ld obeys the
\textit{pseudo-additivity} relation [3].

Defining the \textit{q-deformed} logarithm and the
\textit{q-deformed} exponential as [4]

\begin{equation}
ln _q \left( x \right) = \frac{{x^{1 - q} - 1}}{{1 - q}},
\end{equation}
and
\begin{equation}
e_q^x = \left\{ \begin{array}{l}
 \left[ {1 + \left( {1 - q} \right)x} \right]^{^{{\raise0.7ex\hbox{$1$} \!\mathord{\left/
 {\vphantom {1 {\left( {1 - q} \right)}}}\right.\kern-\nulldelimiterspace}
\!\lower0.7ex\hbox{${\left( {1 - q} \right)}$}}} } ;1 + \left( {1 - q} \right)x \ge 0 \\
 0;otherwise, \\
\end{array} \right.
\end{equation}
 respectively.  The Tsallis entropy (1), and, the generalized K-Ld (3) may be
written as

\begin{equation}
S_q \left( {p\left( x \right)} \right) =  - \sum\limits_x {p\left( x
\right)} ^q \ln _q p\left( x \right),
\end{equation}
and,
\begin{equation}
I_q \left( {p\left( x \right)\left\| {r(x)} \right.} \right) =  -
\sum\limits_x {p\left( x \right)} \ln _q \frac{{r(x)}}{{p(x)}},
\end{equation}
respectively.  Employing the relation [4]
\begin{equation}
\ln _q \left( {\frac{x}{y}} \right) = y^{q - 1} \left( {\ln _q x -
\ln _q y} \right),
\end{equation}
the generalized K-Ld (7) may be expressed as the generalized mutual
information
\begin{equation}
\begin{array}{l}
 I_q\left( {X;\tilde X} \right) =- \sum\limits_{x,\tilde x} {p\left( {x,\tilde x} \right)\ln _q } \left( {\frac{{p\left( {\tilde x} \right)}}{{p\left( {\left. {\tilde x} \right|x} \right)}}}
 \right) \\
= \sum\limits_{\tilde x} {\frac{1}{{p^{q - 1} \left( {\tilde x}
\right)}}} S_q \left( X \right) - \underbrace {\sum\limits_{\tilde
x} {p\left( {\tilde x} \right)} S_q \left( {\left. X \right|\tilde X
= \tilde x} \right)}_{\left\langle {S_q \left( {\left. X
\right|\tilde X = \tilde x} \right)} \right\rangle _{p\left( {\tilde
x} \right)} }. \\
\end{array}
\end{equation}
Here, $ \left\langle  \bullet  \right\rangle $ denotes the
expectation value.  The seminal paper on source coding within the
framework of nonextensive statistics by Landsberg and Vedral [5],
has provided the impetus for a number of investigations into the use
of nonextensive information theory within the context of coding
problems. The works of Yamano [6,7] represent a sample of some of
the prominent efforts in this regard. \emph{The source coding
theorem in a nonextensive setting has been derived by Yamano [8]}.

A statistical physics model for the variational problem encountered
in rate distortion (RD) theory [9, 10] and the information
bottleneck method [11], derived within a generalized statistics
framework, has been recently established [12]. A nonextensive
Blahut-Arimorto (BA) alternate minimization scheme [13] has been
derived. The noteworthy result of the study in [12] is that the
nonextensive RD curves possesses a lower threshold for the minimum
compression information in the distortion-compression plane, as
compared to equivalent RD curves derived on the basis of the
Boltzmann-Gibbs-Shannon framework.  As was the case in [12], this
paper employs values of $ q $ in the range $ 0 < q < 1 $. This paper
utilizes the statistical physics theory presented in [12] to
formulate a generalized Bregman RD (GBRD) model.  The GBRD model
extends previous works by Rose [14] and Banerjee \textit{et. al.}
[15, 16] to achieve a principled and practical strategy to evaluate
RD functions.

\section{Rationale for the GBRD model}

\subsection{Background information}
The RD problem in terms of discrete random variables is stated as
follows: given a discrete random variable $ X \in \Xi $ called the
source or the codebook, and, another discrete random variable $
\tilde X \in \tilde \Xi $ which is a compressed representation of $
X $ (also referred to as the \textit{quantized codebook} and/or the
\textit{reproduction alphabet}), the information rate distortion
function that is to be obtained is the minimization of the mutual
information $ I_{q\geq 1}(X,\tilde X) $ over all conditional
probabilities $ p\left( {\tilde x\left| x \right.} \right) $.

The crux of the RD problem is the numerical determination of the RD
function using the BA scheme. The actual implementation of the BA
scheme is sometimes impractical owing to a lack of knowledge of the
optimal support of the \textit{quantized codebook} $ \tilde X $.
Exact analytical solutions exist only for a few cases consisting of
a combination well behaved sources and distortion measures.  An
initial attempt to achieve a practical and tractable solution to the
RD problem was performed by Rose [14], for the case of Euclidean
square distortion functions. Therein, it was demonstrated that for
sources whose support is a bounded set, the RD function either
equals the \textit{Shannon lower bound}, or, the optimal support for
the \textit{quantized codebook} is finite thus permitting the use of
a numerical procedure called the \textit{mapping method}.  The
pioneering work of Rose was generalized by Banerjee \textit{et. al}
[15, 16] to include a wider class of distortion functions using
Bregman divergences.  One of the significant features in these works
involved the formulation of a \textit{Shannon-Bregman lower bound}.

Motivated by the recent results [12], this paper provides two means
to solve the GBRD problem for sources with bounded support. First,
the analytical solution may be obtained from a
\textit{Tsallis-Bregman lower bound} (Section IV).  Next, for a
finite reproduction alphabet, the RD function may be numerically
obtained by a computational methodology derived from generalized
statistics (Section III).

\subsection{Bregman Divergences}
\textbf{Definition 1 (Bregman divergences)}:  \textit{Let $ \phi $
be a real valued strictly convex function defined on the convex set
$ S \subseteq \Re $, the domain of $ \phi $ such that $ \phi $ is
differentiable on $ int(S) $, the interior of $ S $.  The Bregman
divergence $ B_\phi :S \times {\mathop{\rm int}} \left( S \right)
\mapsto \Re ^ + $ is defined as $ B_\phi  \left( {z_1 ,z_2 } \right)
= \phi \left( {z_1 } \right) - \phi \left( {z_2 } \right) - \left(
{z_1  - z_2 ,\nabla \phi \left( {z_2 } \right)} \right) $, where $
B_\phi  \left( {z_1 ,z_2 } \right) = \phi \left( {z_1 } \right) -
\phi \left( {z_2 } \right) - \left( {z_1  - z_2 ,\nabla \phi \left(
{z_2 } \right)} \right) $ is the gradient of $ \phi $ evaluated at $
z_2 $.}

A number of Bregman divergences have been tabulated in [15]. The
generalized K-Ld in (3), also referred to as the Csisz\'{a}r
generalized K-Ld is not a Bregman divergence.  Since the seminal
work by Naudts [17], Bregman divergences have been the object of
much research in nonextensive statistics.  Defining $ \phi (p) =
 -p\ln _q \left( {{\raise0.7ex\hbox{$1$} \!\mathord{\left/
 {\vphantom {1 p}}\right.\kern-\nulldelimiterspace}
\!\lower0.7ex\hbox{$p$}}} \right) $, the Bregman generalized K-Ld is
defined as
\begin{equation}
\begin{array}{l}
d_{\phi}\left( {\textbf{p},\textbf{r}} \right) = \sum\limits_{i =
1}^d {\frac{{p_i }}{{\left( {q - 1} \right)}}} \left( {p_i^{q - 1} -
r_i^{q - 1} } \right) -
\sum\limits_{i = 1}^d {\left( {p_i  - r_i } \right)} r_i^{q - 1} \\
= I_{q,B} \left( {\left. \textbf{p} \right\|\textbf{r}} \right)
\end{array}
\end{equation}

Setting $ q - 1 \to \kappa $, (10) is consistent with Eq. (35) in
[17].
\section{the GBRD model}
This Section provides a strategy to jointly obtain the optimal
support $ \tilde \mathcal X_s $\footnote{Calligraphic symbols
indicate sets} of the \textit{quantized codebook} with
\textit{cardinality} $ \left| {\tilde \mathcal X_s} \right| = k $ ,
and, the conditional probability $ p(\tilde x|x) $ that
characterizes the RD problem. This is accomplished by a joint
optimization of
\begin{equation}
\mathop {\min }\limits_{\tilde \mathcal X_s ,p\left( {\left. {\tilde
x} \right|x} \right)} L_{GBRD}^q  = \mathop {\min
}\limits_{\scriptstyle \tilde \mathcal X_s,p\left( {\left. {\tilde
x} \right|x} \right) \hfill \atop
  \scriptstyle \left| {\tilde \mathcal X_s} \right| = k \hfill} \left\{ {I_q \left( {X;\tilde X} \right) + \beta \left\langle {d_\phi  \left( {x,\tilde x} \right)} \right\rangle _{p\left( {x,\tilde x} \right)} }
  \right\}.
\end{equation}
The optimal Lagrange multiplier $ \beta $, hereafter referred to as
the \textit{inverse temperature}, depends upon the optimal tolerance
level of the expectation of $ D=<d_{\phi}(x,\tilde x)>_{p(x,\tilde
x)} = \sum\limits_{x,\tilde x} {p\left( {x,\tilde x} \right)} d_\phi
\left( {x,\tilde x} \right) $.
\subsection{Constraint terms}
Generalized statistics has utilized a number of constraints to
define expectations.  The original Tsallis (OT) constraints of the
form $ \left\langle A \right\rangle  = \sum\limits_i {p_i } A_i $
[1], were convenient owing to their similarity to the maximum
entropy constraints.  These were abandoned because of difficulties
encountered in obtaining an acceptable form for the partition
function, with the exception of a few specialized cases.

The OT constraints were subsequently replaced by the Curado-Tsallis
(C-T) constraints $ \left\langle A \right\rangle _q^{C-T}  =
\sum\limits_i {p_i^q } A_i $. The C-T constraints were later
replaced by the normalized Tsallis-Mendes-Plastino (T-M-P)
constraints $ \left\langle A \right\rangle ^{T - M - P}  =
\sum\limits_i {\frac{{p_i^q }}{{\sum\limits_i {p_i^q } }}A_i } $.
The dependence of the expectation value on the normalization
\textit{pdf} $ {\sum\limits_i {p_i^q } } $, renders the T-M-P
constraints to be \textit{self-referential}. This paper, like [12],
utilizes a recent methodology [18] to ``rescue" the OT constraints,
and, has linked the OT, C-T, and, T-M-P constraints.

\subsection{The nonextensive variational principle}
Keeping $ \tilde \mathcal X_s $ fixed, taking the variational
derivative of (11) with respect to $ p(\tilde x|x) $ while enforcing
$ \sum\limits_{\tilde x} {p\left( {\left. {\tilde x} \right|x}
\right)}  = 1 $ with the normalization Lagrange multiplier $
\lambda(x) $, yields
\begin{equation}
\begin{array}{l}
p\left( {\tilde x\left| x \right.} \right) = p\left( {\tilde x}
\right)\left[ {\frac{(1-q)}{{q}}\left\{ {\tilde \lambda \left( x
\right) + \beta d_{\phi}\left( {x,\tilde x} \right)} \right\}}
\right]^{1/\left( {q - 1} \right)},\\
\\
\tilde \lambda \left( x \right) = \frac{{\lambda \left( x
\right)}}{{p\left( x \right)}} - p\left( x \right)^{\left( {1 - q}
\right)}. \\
\end{array}
\end{equation}
Multiplying the terms in the square brackets in (12) by the
conditional probability $ p\left( {\left. {\tilde x} \right|x}
\right) $, and summing over $ \tilde x $ yields
\begin{equation}
\begin{array}{l}
 \frac{q}{{q - 1}}\sum\limits_{\tilde x} {p\left( {\tilde x} \right)\left( {\frac{{p\left( {\left. {\tilde x} \right|x} \right)}}{{p\left( {\tilde x} \right)}}} \right)^q }  + \underbrace {\beta \sum\limits_{\tilde x} {d_{\phi}\left( {x,\tilde x} \right)p\left( {\left. {\tilde x} \right|x} \right)} }_{\beta \left\langle {d_{\phi}\left( {x,\tilde x} \right)} \right\rangle _{p\left( {\left. {\tilde x} \right|X = x} \right)} } + \\
+ \tilde \lambda \left( x \right)\sum\limits_{\tilde x} {p\left( {\left. {\tilde x} \right|x} \right) = 0}. \\
\end{array}
\end{equation}
Evoking $ \sum\limits_{\tilde x} {p\left( {\left. {\tilde x}
\right|X = x} \right)}  = 1 $, yields
\begin{equation}
\tilde \lambda \left( x \right) = \frac{q}{{\left( {1 - q}
\right)}}\aleph _q \left( x \right) - \beta \left\langle
{d_{\phi}\left( {x,\tilde x} \right)} \right\rangle _{p\left(
{\left. {\tilde x} \right|X = x} \right)}.
\end{equation}
Here, $ \aleph _q \left( x \right) = \sum\limits_{\tilde x} {p\left(
{\tilde x} \right)\left( {\frac{{p\left( {\left. {\tilde x}
\right|x} \right)}}{{p\left( {\tilde x} \right)}}} \right)^q } $.
The conditional \textit{pdf} $ p\left( {\left. {\tilde x} \right|x}
\right) $ acquires the form
\begin{equation}
\begin{array}{l}
 p\left( {\left. {\tilde x} \right|x} \right) = \frac{{p\left( {\tilde x} \right)\left\{ {1 - \left( {q - 1} \right)\beta ^ *  d_{\phi}\left( {x,\tilde x} \right)} \right\}^{1/\left( {q - 1} \right)} }}{{\Im_{RD} ^{^{1/\left( {1 - q} \right)} } }}, \\

\Im _{RD}  = q\aleph _q \left( x \right) + \left( {q - 1}
\right)\beta \left\langle {d_{\phi}  \left( {x,\tilde x} \right)}
\right\rangle _{p\left( {\left. {\tilde x} \right|X = x} \right)},\\

\beta ^ *   = \frac{\beta }{\Im _{RD}},
 \end{array}
\end{equation}
where $ \beta ^ * $ is the \textit{effective inverse temperature}.
Transforming $ q \to 2 - q^
* $ in the numerator and evoking (5), (15) is expressed in the form of a \textit{q-deformed}
exponential
\begin{equation}
p\left( {\tilde x\left| x \right.} \right) = \frac{{p\left( {\tilde
x} \right)\exp _{q^ *  }( -\beta^* d_{\phi}\left( {x,\tilde x})
\right)}}{\tilde Z(x,\beta^*) }.
\end{equation}
The partition function evaluated at each instance of the source
distribution is
\begin{equation}
\tilde Z\left( {x,\beta ^ *  } \right) = \Im _{RD}^{\frac{1}{{\left(
{1 - q} \right)}}}  = \sum\limits_{\tilde x} {p\left( {\tilde x}
\right)\exp _q } \left[ { - \beta ^ *  d_{\phi}\left( {x,\tilde x}
\right)} \right]
\end{equation}
The term $ \left\{ {1 - \left( {1 - q^ *  } \right)\beta ^ * d\left(
{x,\tilde x} \right)} \right\} $ in the numerator of (16) is a
manifestation of the \textit{Tsallis cut-off condition} [18].  This
implies that solutions of (16) are valid when $ \beta ^ *  d\left(
{x,\tilde x} \right) < {\raise0.7ex\hbox{$1$} \!\mathord{\left/
 {\vphantom {1 {\left( {1 - q^ *  } \right)}}}\right.\kern-\nulldelimiterspace}
\!\lower0.7ex\hbox{${\left( {1 - q^ *  } \right)}$}} $.  The
\textit{effective nonextensive RD Helmholtz free energy} is
\begin{equation}
F_{RD}^q \left( {\beta ^ *  } \right) = \frac{{ - 1}}{{\beta ^ *
}}\left\langle {\ln _q \tilde Z\left( {x,\beta ^ *  } \right)}
\right\rangle _{p\left( x \right)}  = \frac{1}{{\beta ^ *
}}\left\langle {\frac{{\Im_{RD}  - 1}}{{q - 1}}} \right\rangle
_{p\left( x \right).}
\end{equation}

Solution of (16) may be viewed from two distinct perspectives, i.e.
the \textit{canonical perspective} and the \textit{parametric
perspective}. Owing to the \textit{self-referential} nature of the
\textit{effective inverse temperature} $ \beta^* $, the analysis and
solution of (16) within the context of the \textit{canonical
perspective} is a formidable undertaking. For practical
applications, the parametric perspective is utilized by evaluating
the conditional \textit{pdf} $ p\left( {\left. {\tilde x} \right|x}
\right) $, employing the nonextensive BA algorithm, for
\textit{a-priori} specified $ \beta ^
*   \in \left[ {0,\infty } \right] $.  \textit{Note that within the context
of the parametric perspective, the self-referential nature of $
\beta^* $ vanishes}. The \textit{inverse temperature} $ \beta $ and
the \textit{effective inverse temperature} $ \beta^* $ relate as
\begin{equation}
\beta  = \frac{{q\aleph _q \left( x \right)\beta ^ *  }}{{\left[
1-{\beta ^ *  \left( {q - 1} \right)\left\langle {d_{\phi}\left(
{x,\tilde x} \right)} \right\rangle _{p\left( {\left. {\tilde x}
\right|X = x} \right)} } \right]}}.
\end{equation}
\subsection{Support estimation step}
The procedure in Section III.B. is carried out for a given $ \beta
$, corresponding to a single point on the RD curve.  Reproduction
alphabets with optimal support are obtained by keeping $ p(\tilde
x|x) $ fixed, and solving
\begin{equation}
\mathop {\min }\limits_{\tilde \mathcal X_s } L_{GBRD}^q  = \mathop
{\min }\limits_{\scriptstyle \tilde \mathcal X_s} \left\{ {I_q
\left( {X;\tilde X} \right) + \beta \left\langle {d_\phi  \left(
{x,\tilde x} \right)} \right\rangle _{p\left( {x,\tilde x} \right)}
} \right\}.
\end{equation}
The solution to (20) is the optimal estimate/predictor [15, 16]
\begin{equation}
\tilde x^ *   = \left\langle x \right\rangle _{p\left( {\left. x
\right|\tilde X = \tilde x} \right)} = \sum\limits_x {p\left(
{\left. x \right|\tilde X = \tilde x} \right)x}.
\end{equation}
Algorithm 1 depicts the \textit{pseudo-code} for the greedy joint
non-convex optimization procedure that constitutes the solution to
the GBRD model.
\section{the Tsallis-Bregman lower bound}
\textbf{Theorem 1} \textit{For OT constraints, the nonextensive RD
function with source $ X \sim p(x) $ and a Bregman divergence $
d_{\phi} $ is always lower bounded by the Tsallis-Bregman lower
bound defined by}
\begin{equation}
\begin{array}{l}
R_{qL} \left( D \right) = \sum\limits_{\tilde x}
{\frac{1}{{p^{q-1}\left( {\tilde x} \right)}}} S_q \left( X \right)
+ \\
+ \mathop {\sup }\limits_{\beta  \ge 0} \left\{ { - \beta D +
\left\langle {\ln _{q^
*  } \gamma _{L,\beta^* } \left( x \right)} \right\rangle _{p\left( x
\right)} } \right\},
\end{array}
\end{equation}
\textit{where $ \gamma _{L\beta ^ *  } $ is a unique function
satisfying}
\begin{equation}
\int_{dom(\phi)} {p\left( t \right)\gamma _{L,\beta ^ *  } \left( t
\right)\exp _{q^ *  } \left( { - \beta ^ *  d_{\phi}\left( {t,\mu}
\right)} \right)dt}  = 1; \forall \mu  \in dom\left( \phi  \right).
\end{equation}
To highlight the critical dependence of the \textit{Tsallis-Bregman
lower bound} upon the type of constraint employed to solve the GBRD
variational problem in Section III B, and enunciate certain aspects
of \textit{q-deformed} algebra and calculus, a slightly weaker bound
(see Appendix C, [16]) is stated and proved.

\textbf{Theorem 2}  \emph{The generalized RD function is defined by
\begin{equation}
R_q \left( D \right) = \mathop {\sup }\limits_{\beta  \ge 0,\gamma^*
\in \Lambda _{\beta^*}  } \left\{ { - \beta D + \int_x {p\left( x
\right)\ln _{q^ *  } \gamma^* \left( x \right)dx} } \right\} ,
\end{equation}
where $ {\Lambda _{\beta ^ *  } } $ is the set of all admissible
partition functions.  Further, for each $ {\beta  \ge 0} $, a
necessary and sufficient condition for $  \gamma^* \left( x \right)
$ to attain the supremum in (24)is a probability density $ p(\tilde
x) $ related to $ \gamma^*(x) $ as
\begin{equation}
\gamma^* \left( x \right) = \left( {\int_{\tilde x} {p\left( {\tilde
x} \right)} \exp _q \left( { - \beta ^ *  d_\phi \left( {x,\tilde x}
\right)} \right)d\tilde x} \right)^{ - 1}.
\end{equation}}
\textbf{Proof}
\begin{equation}
\begin{array}{l}
 L_{GBRD}^q \left[ p({\left. {\tilde x} \right|x} \right)] =  \\
  - \int_{x,\tilde x} {p\left( x \right)} p\left( {\left. {\tilde x} \right|x} \right)\ln _q \left( {\frac{{p\left( {\tilde x} \right)}}{{p\left( {\left. {\tilde x} \right|x} \right)}}} \right)dxd\tilde x +  \\
  + \beta \int_{x,\tilde x} {p\left( x \right)} p\left( {\left. {\tilde x} \right|x} \right)d_\phi  \left( {x,\tilde x} \right)dxd\tilde x \\

 \mathop  = \limits^{\left( a \right)} \int_{x,\tilde x} {p\left( x \right)} p\left( {\left. {\tilde x} \right|x} \right)\ln _{q^ *  } \left( {\frac{{p\left( {\left. {\tilde x} \right|x} \right)}}{{p\left( {\tilde x} \right)}}} \right)dxd\tilde x +  \\
  + \beta \int_{x,\tilde x} {p\left( x \right)} p\left( {\left. {\tilde x} \right|x} \right)d_\phi  \left( {x,\tilde x} \right)dxd\tilde x \\

 \mathop  = \limits^{\left( b \right)} \int_{x,\tilde x} {p\left( x \right)} p\left( {\left. {\tilde x} \right|x} \right)\ln _{q^ *  } \left( {\frac{{\exp _{q^ *  } \left( { - \beta ^ *  d_\phi  \left( {x,\tilde x} \right)} \right)}}{{\tilde Z\left( {x,\beta ^ *  } \right)}}} \right)dxd\tilde x +  \\
  + \beta \int_{x,\tilde x} {p\left( x \right)} p\left( {\left. {\tilde x} \right|x} \right)d_\phi  \left( {x,\tilde x} \right)dxd\tilde x \\

 \mathop  = \limits^{\left( c \right)}  - \int_{x,\tilde x} {p\left( x \right)p\left( {\left. {\tilde x} \right|x} \right)} \tilde Z^{\left( {q - 1} \right)} \left( {x,\beta ^ *  } \right)\left( {\beta ^ *  d_\phi  \left( {x,\tilde x} \right) + } \right. \\
 \left. {\ln _{q^ *  } \tilde Z\left( {x,\beta ^ *  } \right)} \right)dxd\tilde x + \beta\int_{x,\tilde x} {p\left( x \right)p\left( {\left. {\tilde x} \right|x} \right)} d_\phi  \left( {x,\tilde x} \right)dxd\tilde x \\

 \mathop  = \limits^{\left( d \right)}  - \int_{x,\tilde x} {p\left( x \right)p\left( {\left. {\tilde x} \right|x} \right)} \ln _q \tilde Z\left( {x,\beta ^ *  } \right)dxd\tilde
 x \\

  = - \int_x {p\left( x \right)} \ln _q \tilde Z\left( x,\beta ^ * \right)dx \\

 \mathop  = \limits^{\left( e \right)}  - \int_x {p\left( x \right)} \ln _q \left( {\int_{\tilde x} {p\left( {\tilde x} \right)\exp _{q^ *  } \left( { - \beta ^ *  d_\phi  \left( {x,\tilde x} \right)} \right)d\tilde x} } \right)dx \\

=L_{GBRD}^q \left[ p(\tilde x \right)].
\end{array}
\end{equation}
Here, $ (a) $ follows from $ \ln _q \left( x \right) =  - \ln _{q^
*  } \left( {{\raise0.7ex\hbox{$1$} \!\mathord{\left/
 {\vphantom {1 x}}\right.\kern-\nulldelimiterspace}
\!\lower0.7ex\hbox{$x$}}} \right);q^ *   = 2 - q $, $ (b) $ follows
from (16) where the partition function in the continuum is $ \tilde
Z\left( {x,\beta ^ *  } \right) = \int_{\tilde x} {p\left( {\tilde
x} \right)} \exp _q \left( { - \beta ^ *  d_\phi  \left( {x,\tilde
x} \right)} \right)d\tilde x $, $ (c) $ follows from (8), $ (d) $
follows from (15), and, $ (e) $ follows from (17). \textit{Analogous
to the \textit{mapping method} of [14], a mapping from the unit
interval to the reproduction alphabet is sought such that the
Lebesgue measure over the unit interval results in an optimal $
p(\tilde x) $ over the reproduction alphabet}.  Thus, for each
instance of a probability measure corresponding to $ p\left( {\tilde
x} \right) $ on $ \tilde \mathcal X $ there exists a unique map $
\tilde x:\left[ {0,1} \right] \mapsto \tilde \mathcal X $ that maps
the Lebesgue measure ($ \mu $) to $ p\left( {\tilde x} \right) $
such that for any function $ f $ defined over $ \tilde x $, $
\int_{\tilde x} {f\left( {\tilde x} \right)p\left( {\tilde x}
\right)} d\tilde x = \int_{u \in \left[ {0,1} \right]} {f\left(
{\tilde x\left( u \right)} \right)} d\mu \left( u \right) $.  Thus,
(26) yields
\begin{equation}
\begin{array}{l}
 L_{GBRD}^q \left[ {\tilde x} \right] = \\

 = - \int_x {p\left( x \right)} \ln _q \left( {\int_{\tilde x} {p\left( {\tilde x} \right)\exp _{q^ *  } \left( { - \beta ^ *  d_\phi  \left( {x,\tilde x} \right)} \right)d\tilde x} } \right)dx, \\

  =  - \int_x {p\left( x \right)} \ln _q \left( {\int_{u \in \left[ {0,1} \right]} {\exp _{q^ *  } \left( { - \beta ^ *  d_\phi  \left( {x,\tilde x\left( u \right)} \right)} \right)d\mu \left( u \right)} } \right)dx .\\
 \end{array}
\end{equation}

To obtain the optimality condition, the \textit{q-deformed} terms
characteristic to generalized statistics are to be operated on by
the \textit{dual Jackson derivative} instead of the usual Newtonian
derivative [4].  To obtain expressions analogous to the case of
Newton-Leibniz calculus, the \textit{dual Jackson derivative}
defined as
\begin{equation}
\begin{array}{l}
 D^{\left( q \right)} f\left( x \right) = \frac{1}{{1 + (1 - q)f\left( x \right)}}\frac{{df}}{{dx}};1 + (1 - q)f\left( x \right) \ne 0 \\
  \Rightarrow D^{\left( q \right)} \ln _q \tilde Z\left( {x,\beta ^ *  } \right) =  \frac{1}{{\tilde Z\left( {x,\beta ^ *  } \right)}}\frac{{d\tilde Z\left( {x,\beta ^ *  } \right)}}{{dx}}
\end{array}
\end{equation}
is employed to enforce the optimality condition.   The optimality
condition for (27) becomes
\begin{equation}
D^{\left( q \right)} L_{GBRD}^q  = -\int_x {\frac{{p\left( x
\right)\frac{d}{{d\tilde x}}\left( {\exp _{q^*} \left( { - \beta ^ *
d_\phi  \left( {x,\tilde x\left( u \right)} \right)} \right)}
\right)dx}}{{\int_{u \in \left[ {0,1} \right]} {\exp _{q^*} \left( {
- \beta ^ *  d_\phi  \left( {x,\tilde x\left( u \right)} \right)}
\right)ud\mu } }}}.
\end{equation}
Evoking \textit{Borel isomorphism}, and, assuming that the optimal
support $ \tilde \mathcal X_s $ contains a non-empty open ball $
B_{\epsilon} $, $ \epsilon > 0 $ implies that
\begin{equation}
\begin{array}{l}
D^{\left( q \right)} L_{GBRD}^{\left( q \right)}  = \int_x
{\frac{{p\left( x \right)\exp _{q^ *  } \left( { - \beta ^ * d\left(
{x,\tilde x} \right)} \right)dx}}{{N\left( {x,\beta ^ *  }
\right)}}}  = k_o \forall \tilde x \in B_{\epsilon} \subset \tilde
\mathcal X_s, \\
\Rightarrow \int_{\tilde x \in B_ \in  } {\int_x {\frac{{p\left( x \right)p\left( {\tilde x} \right)\exp _{q^ *  } \left( { - \beta ^ *  d\left( {x,\tilde x} \right)} \right)dxd\tilde x}}{{N\left( {x,\beta ^ *  } \right)}}} }  = k_o \int_{\tilde x \in B_ \in  } {p\left( {\tilde x} \right)d\tilde x} , \\
  \Rightarrow \int_{\tilde x \in B_ \in  } {\int_x {p\left( x \right)p\left( {\left. {\tilde x} \right|x} \right)dxd\tilde x = } } k_o \int_{\tilde x \in B_ \in  } {p\left( {\tilde x} \right)d\tilde x} \Rightarrow 1 = k_o,  \\
 \end{array}
\end{equation}
holds true for all $ \tilde x \in B_{\epsilon} $. Defining $ \gamma
_{\beta ^ *  }^ *  \left( x \right) = \left( {\int_{\tilde x}
{p\left( {\tilde x} \right)} \exp _q \left( { - \beta ^ *  d_\phi
\left( {x,\tilde x} \right)} \right)d\tilde x} \right)^{ - 1} $,
(30) yields
\begin{equation}
\int_x {p\left( x \right)\gamma _{\beta ^ *  }^ *  \left( x
\right)\exp _{q^ *  } \left( { - \beta ^ *  d\left( {x,\tilde x}
\right)} \right)dx}  = 1;\forall \tilde x \in B_ \in.
\end{equation}
Using the relation $ - \ln _q \left( x \right) = \ln _{q^*} \left(
{{\raise0.7ex\hbox{$1$} \!\mathord{\left/
 {\vphantom {1 x}}\right.\kern-\nulldelimiterspace}
\!\lower0.7ex\hbox{$x$}}} \right) $, (26) becomes
\begin{equation}
L_{GBRD}^q \left[ {p\left( {\tilde x} \right)} \right] = \int_x
{p\left( x \right)\ln _{q^ *  } \gamma _{\beta ^ *  }^ *  \left( x
\right)} dx.
\end{equation}
Thus, $ \gamma _{\beta ^* }^
*(x) $ satisfies (25) and attains the supremum in (24) for a given $
\beta $ and corresponding $ \beta ^ * $
\begin{equation}
R_q \left( {D_\beta  } \right) =  - \beta D_\beta   + \int_x
{p\left( x \right)\ln _{q^ *  } \gamma _{\beta ^ *  }^ *  \left( x
\right)dx},
\end{equation}
where $ D_{\beta} $ is the distortion value at which the supremum in
(24) is attained for a given $ \beta $.  Thus, the
\textit{Tsallis-Bregman lower bound }is
\begin{equation}
R_{qL} \left( {D_\beta  } \right) =  - \beta D_\beta   + \int_x
{p\left( x \right)\ln _{q^ *  } \gamma _{\beta ^ *  }^ *  \left( x
\right)dx}=R_{q} \left( {D_\beta  } \right).
\end{equation}

\begin{algorithm}
\caption{GBRD Model}
\begin{algorithmic}
\STATE \textbf{Input} \\

1. $ X \sim p(x) $ over $ \left\{ {x_i } \right\}_{i = 1}^n  \subset
dom\left( \phi  \right) \subseteq \Re ^m $.\\
2. Bregman divergence $ d_{\phi} $, $ | \tilde \mathcal X_s |=k $,
\textit{effective variational parameter} $ \beta ^ *   \in \left[
{0,\infty } \right] $, each $ \beta ^ * $ is a single point on the
RD curve.

\STATE \textbf{Output} \\
1.$ \tilde \mathcal X_s^ *   = \left\{ {\tilde x} \right\}_{j = 1}^k
,P^
* = \left\{ {\left\{ {\left. {\tilde x_j } \right|x_i } \right\}_{j
= 1}^k } \right\}_{i = 1}^n $ that locally optimizes (11), $
(R_{\beta},D_{\beta}) $ tradeoff at each $ \beta^* $ \\
2.Value of $ R_q(D) $ where its slope equals $ -\beta  =
\frac{{-q\aleph _q \left( x \right)\beta ^ * }}{{\left[ 1-{\beta ^
* \left( {q - 1} \right)\left\langle {d_{\phi}\left( {x,\tilde x} \right)}
\right\rangle _{p\left( {\left. {\tilde x} \right|X = x} \right)} }
\right]}}. $

\STATE \textbf{Method} \\

1.Initialize with some $ \left\{ {\tilde x} \right\}_{j = 1}^k
\subset dom\left( \phi  \right) $. \\
2.Set up outer $ \beta^* $ loop. \\
\textbf{repeat}\\
{Blahut-Arimoto loop (16) for single value of $ \beta^* $}\\
\textbf{repeat} \\
for i=1 to $ n $ do \\
     for j=1 to $ k $ do \\
$ p\left( {\tilde x_j\left| x_i \right.} \right) \leftarrow
\frac{{p\left( {\tilde x_j} \right)\exp _{q^ *  }( -\beta^*
d_{\phi}\left( {x_i,\tilde x_j}) \right)}}{\tilde Z(x_i,\beta^*) } $ \\
end for \\
end for \\
for j=1 to $ k $ do \\

$ p\left( {\tilde x_j } \right) \leftarrow \sum\nolimits_{i = 1}^n
{p\left( {\left. {\tilde x_j } \right|x_i } \right)p\left( {x_i }
\right)} $ \\
end for \\
until \textit{convergence} \\
{Support Estimation Step ($ \tilde \mathcal X_s $ using (21))} \\
for j=1 to $ k $ do \\
$ \tilde x_j  \leftarrow \sum\nolimits_{i = 1}^n {p\left( {\left.
{x_i } \right|\tilde x_j } \right)} x_i $ \\
\textbf{end for} \\
\textbf{until} \textit{convergence} \\
Calculate $ D_{\beta}  $ and $ R_{\beta} $ for the $ \beta^* $.  \\
\textbf{advance} $ \beta^* $ \\
$ \beta ^ *   = \beta ^ *   + \delta \beta ^ * $
\end{algorithmic}
\end{algorithm}

\section{Numerical simulations}
The efficacy of the GBRD model is computationally investigated by
drawing a sample of 1000 two-dimensional data points, from three
spherical Gaussian distributions with centers $ (2, 3.5),(0,
0),(0,2) $ (the \textit{quantized codebook}). The priors and
standard deviations are $ 0.3,0.4,0.3 $, and, $ 0.2,0.5,1.0 $,
respectively.  \textit{To test effectiveness of the \textit{support
estimation step}, the \textit{quantized codebook} is shifted from
the true means to positions at the edges of the spherical Gaussian
distributions}.  The computational procedure described in Algorithm
1 is repeatedly solved for each value of $ \beta^* $, till a
reproduction alphabet with optimal support is obtained. This
consistently coincides with the true mean, for a negligible error.
This effect is particularly pronounced, and rapidly achieved, for
regions of low and intermediate values of $ \beta^* $, thus
providing implicit proof of the relation between soft clustering and
RD with Bregman divergences [15].

Fig.1 depicts the RD curves for extensive RD with Bregman
divergences [15, 16] and the GBRD model, with the constituent
discrete points overlaid upon them. A Euclidean square distortion (a
Bregman divergence) is employed.  Each curve has been generated for
values of $ \beta \in \left[ {.1,2.5} \right] $ (the extensive
case), and $ \beta ^ * \in \left[ {.1,100} \right] $ (the GBRD
cases), respectively. \textit{Note that for the GBRD cases, the
slope of the RD curve is $ -\beta $ and not $ -\beta^* $}.
\textit{Note that all GBRD curves inhabit the non-achievable (no
compression) region of the extensive RD model with Bregman
divergences}. \textit{Further, GBRD models possessing a lower
nonextensivity parameter $ q $ inhabit the non-achievable regions of
GBRD models possessing a higher value of $ q $}.  It is observed
that the GBRD model undergoes compression and clustering more
rapidly than the equivalent extensive RD model with Bregman
divergences. A primary cause for such behavior is the rapid increase
in $ \beta^* $ for marginal increases in $ \beta $, as depicted in
Fig. 2 and obtained from (19).

\section*{Acknowledgement}

This work was supported by \textit{RAND-MSR contract} \textit{CSM-DI
$ \ \& $ S-QIT-101107-2005}.  Gratitude is expressed to A. Plastino,
S. Abe, and, K. Rose for helpful discussions.



%

\begin{figure}[thpb]
\centering
\includegraphics[scale=0.42]{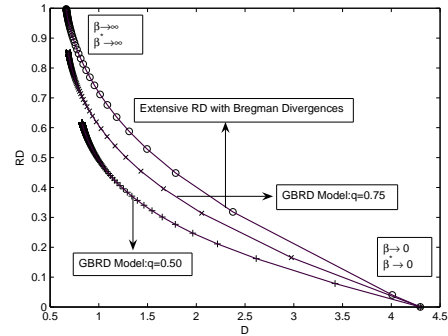}
\caption{Rate Distortion Curves for GBRD Model and Extensive RD with
Bregman Divergences}
\end{figure}

\begin{figure}[thpb]
\centering
\includegraphics[scale=0.42]{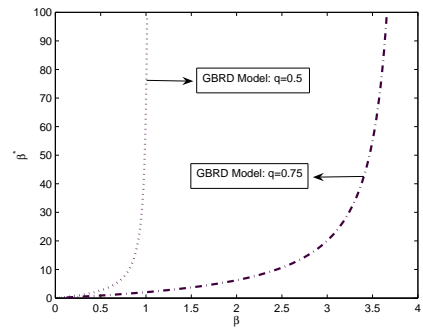}
\caption{Curves for $ \beta $ v/s $ \beta^* $ for GBRD Model}
\end{figure}
\end{document}